\documentclass[prd,nofootinbib,showpacs,showkeys]{revtex4}

\usepackage{hyperref}
\usepackage{epic,amsmath,graphicx}
\usepackage{axodraw,pstricks}
\begin{document}
\preprint{TKU-08-02}

\title{Is the $Z^{+}(4430)$ a radially excited state of $D_s$?}

\author{Takayuki Matsuki}
\email[E-mail: ]{matsuki@tokyo-kasei.ac.jp}
\affiliation{Tokyo Kasei University,
1-18-1 Kaga, Itabashi, Tokyo 173-8602, JAPAN}
\author{Toshiyuki Morii}
\email[E-mail: ]{morii@kobe-u.ac.jp}
\affiliation{Graduate School of Human Development and Environment,
Kobe University, 3-11 Tsurukabuto, Nada, Kobe 657-8501, JAPAN}
\author{Kazutaka Sudoh}
\email[E-mail: ]{kazutaka.sudoh@kek.jp}
\affiliation{Nishogakusha University,
6-16 Sanbancho, Chiyoda, Tokyo 102-8336, JAPAN}
%

\date{August 15, 2008} %
\begin{abstract}
We present the interpretation that the recently discovered $Z^{+}(4430)$ by the Belle Collaboration can be a radial excitation of the $c\bar s$ state, being consistent with an observed value of the product of branching ratios, ${\cal B}(B^0\to K^\mp Z^\pm(4430))\times {\cal B}(Z^\pm(4430)\to \pi^\pm\psi')\sim 10^{-5}$.
We give an explicit $c\bar s$ candidate for this state by calculating the mass 
value in our semirelativistic  quark potential model and also give a natural 
understanding for the facts that the decay mode $Z\to J/\psi \pi^+$ has not yet been seen 
while $Z\to \psi' \pi$ can be seen.
\end{abstract}
\maketitle

A series of exotic $X$, $Y$, and $Z$ charmonium-like mesons have been discovered by the B 
factories, among which the recent discovery of $Z^{+}(4430)$ by the Belle 
Collaboration 
\cite{BaBar} draws attention of many physicists because of the following 
reasons; $Z^{+}(4430)$ is charged and hence cannot be a $c\bar c$ charmonium 
state or a $c\bar c g$ hybrid meson and it might be the first charged tetraquark state because it is too heavy to be a charged $Q\bar q$ meson and furthermore, strangely enough, only the decay mode 
$Z^{+}\to \psi' \pi^+$ is found while the mode $Z^{+}\to J/\psi \pi^+$ has not 
yet been seen. The product of branching ratios is determined to be 
\begin{equation}
  {\cal B}(B^0\to K^\mp Z^\pm(4430))\times {\cal B}(Z^\pm(4430)\to \pi^\pm\psi')
  =(4.1 \pm 1.0({\rm stat})\pm 1.4({\rm syst}))\times 10^{-5}.
  \label{prdrate}
\end{equation}
Mass and width of this particle are measured as
\begin{equation}
  m_Z=(4433\pm 4 \pm 2) {\rm ~MeV}, \quad
  \Gamma _Z  = \left( {45\begin{array}{*{20}c}
   { + 18 + 30}  \\
   { - 13 - 13}  \\
  \end{array}} \right)
 {\rm ~MeV}.
 \label{masswidth}
\end{equation}

Looking at the above data, many people regard this particle $Z^{+}(4430)$ as 
a strong and plausible candidate of a tetraquark state.  According to such a point of view, the second factor of branching ratios given by Eq.~(\ref{prdrate}) is in the ${\cal O}(1)$, because the decay width of $Z^{+}(4430)$ is 45~MeV as given by  Eq.~(\ref{masswidth}) and the decay $Z^\pm(4430)\to \pi^\pm\psi'$ occurs through strong interaction, and then the first factor becomes the ${\cal O}(10^{-5})$ from Eq.~(\ref{prdrate}), whose suppression is due to a very small 
recombination probability for making four quarks ($c\bar cq\bar q$) into a 
tetraquark state.
%
%
There already appear several papers \cite{Rosner}-\cite{Li} on this 
state whether a tetraquark or molecular state interpretation is possible or not.  However, all models presented so far cannot give a convincing explanation on why the mode $Z^{+}\rightarrow J/\psi\pi^{+}$ has not been seen, though some people \cite{Meng} partly account for the suppression of 
$Z^{+}\rightarrow J/\psi\pi^{+}$ based on the $D_1D^{*}$ or $D_1'D^{*}$ 
resonance model of $Z^{+}(4430)$. Furthermore, it should be noted that the decay width given by Eq.~(\ref{prdrate}) is not a partial decay width for the process $Z^{+}\rightarrow J/\psi\pi^{+}$ but a total decay width of $Z^{+}(4430)$, because it is obtained from fitting the invariant mass distribution of $\pi^{+}+\psi'$ to the Breit-Wigner resonance formula. 
%
%

In this letter, we would like to propose a different interpretation from these papers, i.e., this
state can be a higher radial excitation of $D_s$ state.  In this model, the first factor,
${\cal B}(B^0\to K^\mp D_s^\pm(4430))$, of the branching ratios of Eq.~(\ref{prdrate}) replacing
$Z^{\pm}(4430)$ with $D_s^{\pm}(4430)$ is of the order ${\cal O}(1)$ because the numerator
contains a gluon interaction as shown in Fig.1 below, while the second factor,
${\cal B}(D_s^\pm(4430)\to \pi^\pm\psi')$, is of the order ${\cal O}(10^{-5})$ because an excited
$D_s^{+}(4430)$
can decay to $D^{0}K^{+}/D^{+}K^{0}$ and also to their excited particles $D^{(*)}K^{(*)}$
via strong interactions with the decay width of tens of MeV
because of its large mass value 4430 MeV, in addition to
the decay $D_s^{+}(4430)\to \psi' + \pi^{+}$ occurring via weak interactions. 
%
%
Based on our semirelativistic quark
potential model \cite{MM97}-\cite{MS07}, we give not only the numerical mass value corresponding to $Z^{+}(4430)$ but also give the reason why the decay mode
$Z^{+}\to J/\psi \pi^+$ has not yet been seen. 

The masses of radially excited $D_s(c\bar s)$ states are calculated and presented in Table
\ref{masslevel} by
using our semirelativistic quark potential model, which succeeds in reproducing the mass levels of all existent
heavy mesons including recently discovered higher states of $D, D_s, B$ and $B_s$ \cite{MMS07, MMS072}. In Table
\ref{masslevel}, the same physical parameters as in Refs. \cite{MMS07, MMS072, MMS073} are used and given in
Table \ref{parameter} except for $\alpha_s^{n=i}$ where $i=1, 2, 3,$ and $4$ denote the principal quantum number.
In this paper we have adopted
the same value  $\alpha_s^i$ for $i=3$ and 4 as that for $i=2$ \cite{MMS072}.
From Table \ref{masslevel},
one may identify one of two states,
$n=4~^3P_1(1^{+})~4440$ MeV, and $n=4~^3P_2(2^{+})~4411$ MeV, as a candidate for $Z^{+}(4430)$.
Here we have discarded the last two columns of Table \ref{masslevel} because these states,
$^3D_1(1^{-})$ and $"^3D_2(2^{-})"$ in each row, are largely affected by the off-diagonal elements of
the interaction Hamiltonian among these states with the quantum number $^{2S+1}L_J$ in the larger state
space \cite{MM97}.
Furthermore, it is natural to consider that the $Z^+(4430)$ does not have a large orbital 
angular momentum but that it is rather the $S$ state or at most the $P$ state, even if it has large $n$.

The parity is not a good 
quantum number in the decay process $D_s(4430) \to \psi' \pi^{+}$ since this 
process goes via a weak interaction as shown in Fig. 3(a) and furthermore the 
spin of the excited $D_s$ is 1 if the particles
$\psi'$ and $\pi^{+}$ originated from the excited $D_s$ are in the relative $S$ state, while it can be 2, 1 or 0 if
they are in the relative $P$ state.   Thus, both of $n=4$ $"^3P_1"(1^{+})$ and  $n=4$ $^3P_2(2^{+})$ can be a possible candidate.  This should be compared with the case of $Z^{+}(4430)$ to be a tetraquark, in
which a decay of tetraquark to $\psi'$+$\pi$ occurs through a strong interaction and hence the allowed spin-parity
of a tetraquark should be $1^{+}$ for the relative $S$ state of $\psi'$ and $\pi^{+}$ or $2^{-}, 1^{-},$ and
$0^{-}$ for the relative $P$ state of $\psi'$ and $\pi^{+}$ because the parity is conserved in a strong
interaction process.

\begin{table}[t!]
\caption{First order mass spectra of the radial excited $D_s$ ($n=2,3,4$) in $1/m_Q$. Units are in MeV. }
\label{masslevel}
\begin{tabular}{lcccccccc}
\hline
\hline
$n=2$
& ~~$^1S_0(0^-)$ & ~~$^3S_1(1^-)$ & ~$^3P_0(0^+)$ & ~$"^3P_1"(1^+)$
& ~~$"^1P_1"(1^+)$ & ~~$^3P_2(2^+)$ & ~$^3D_1(1^-)$ & ~$"^3D_2"(2^-)$ \\
\hline
observed
& ~~-- & ~~2715 & ~2856 & ~~--
& ~~-- & ~~--   & ~~--  & ~~-- \\
calculated
& ~~2563 & ~~2755 & ~2837 & ~3082
& ~~3094 & ~~3157 & ~~4449 & ~~1366 \\
\hline
$n=3$
&  &  &  & 
&  &  &  &  \\
\hline
observed
& ~~-- & ~~-- & ~~-- & ~~--
& ~~-- & ~~-- & ~~-- & ~~-- \\
calculated
& ~~3214 & ~~3371 & ~~3488 & ~~3792
& ~~3670 & ~~3736 & ~~3356 & ~~3413 \\
\hline
$n=4$
&  &  &  & 
&  &  &  &  \\
\hline
observed
& ~~-- & ~~-- & ~~-- & ~~--
& ~~-- & ~~-- & ~~-- & ~~-- \\
calculated
& ~~3763 & ~~3870 & ~~4074 & ~~4440
& ~~4333 & ~~4411 & ~~3906 & ~~3966 \\
\hline
\hline
\end{tabular}
\end{table}
\begin{table}[t]
\caption{Optimal values of parameters.}
\label{parameter}
\begin{tabular}{lccccccc}
\hline
\hline
Parameters 
& ~~$\alpha_s^{n=2,3,4}$ & ~~$a$ (GeV$^{-1}$) 
& ~~$b$ (GeV) & ~~$m_{u, d}$ (GeV) & ~~$m_s$ (GeV) & ~~$m_c$ (GeV) & ~~$m_b$ (GeV) \\
& ~~0.344 & ~~1.939 & ~~0.0749 
& ~~0.0112 & ~~0.0929 & ~~1.032 & ~~4.639 \\
\hline
\hline
\end{tabular}
\end{table}

Next, if the $Z^{+}(4430)$ is a radially excited state of $D_s$, how can we 
explain the suppression of the mode $Z^{+}\to J/\psi\pi$?  The answer is as 
follows. Let us consider the decay process $D_s^{+}(4430)\to \psi' \pi$ or 
$D_s^{+}(4430)\to J/\psi \pi$ as $(c\bar s)\to (c\bar c)+\pi$ by treating 
the $\pi$ as an elementary Nambu-Goldstone boson.  Then the decay amplitude is 
proportional to the overlapping integral of the wave functions of $(c\bar s)$ 
and $(c\bar c)$ states.  We notice that if the node of an initial $c\bar s$ 
state wave function is the same as that of a final $c\bar c$ state wave 
function, then the decay amplitude is expected to be large. On the other hand, 
if they are different, the magnitude of the decay amplitude is small.  Here, we assume that the initial state is a radially excited $D_s$ state with higher 
node.  Then, if the 
final state is $J/\psi$ whose node is zero, the magnitude of the decay amplitude is small or negligible.  To see how it works, let us assume the trial wave 
functions for $D_s$, $J/\psi$, and $\psi'$ being expressed by the Hermite 
polynomials, $\Psi_{X}(x,y,z)$, as 
\begin{equation}
  \Psi_{X}(x,y,z)= \frac{1}{N_0} H_{2m}(m_X~ x) H_{2m}(m_X~ y) H_{2m}(m_X~ z)
  e^{-m_X^2(x^2+y^2+z^2)/2}
  \label{wavef},
\end{equation}
where $H_{2m}(x)$ is the $2m$-th Hermite polynomial with a node $m$ and we 
have confined the wave function in the range 
$0\le x <\infty$, $0\le y <\infty$, and $0\le z <\infty$.
$N_0$ is a normalization and $m_X$ is mass of the particle $X$, which is introduced to make the arguments of the
Hermite polynomials dimensionless.  The ratio of decay rates of
$D_s(4430)\to J/\psi \pi$ to $D_s(4430)\to \psi' \pi$ is proportional to  
$|\int d^3x\Psi_{Xm} \Psi_{X0}|^2/|\int d^3x\Psi_{Xm} \Psi_{X1}|^2$ where $m$,
0, and 1 are nodes of the $D_s(4430)$, $J/\psi$, and $\psi'$, respectively. The calculated result is roughly
given by 
\begin{equation}
  \frac{{\Gamma (D_s(4430)  \to J/\psi \pi )}}{{\Gamma (D_s(4430) \to \psi '\pi )}} =
  \left\{ {\begin{array}{*{20}l}
   {6.7 \times 10^{ - 4} } \quad (n=2{\rm ~for~} D_s) \\
   {8.5 \times 10^{ - 6} } \quad (n=3) \\
   {6.9 \times 10^{ - 7} } \quad (n=4) \\
  \end{array}} \right. \label{rareratio} ,
\end{equation}
where $n$ is the principal quantum number of the initial $D_s(4430)$ state and 
its node number is given by $n-1$. We know that the node of $J/\psi$ is zero and that of $\psi'$ is one, hence if we assume the node of $D_s(4430)$ is three 
with $n=4$,
then, the ratio of the production rate for the mode $Z\to J/\psi \pi$ to $Z\to \psi' \pi$ becomes $6.9\times 10^{-7}$, which is negligibly small.

Some comments are in order for distinguishing the cases of $Z={\rm tetraquark}$ 
and $Z={\rm radially~excited}~D_s$.  The chain decay process
$\bar B^0 \to Z^+(4430)K^- \to \psi' \pi^+ K^-$ for $Z^{+}(4430)$ being a radially excited $D_s$ or a 
tetraquark goes through the Feynman diagrams shown in Fig. 1 and Fig. 2,
respectively. As mentioned early, the excited $D_s(4430)$ decays via strong interactions as $D_s(4430)\to DK$ ($D_s^+(4430)\to D^+k^0/D^0K^+$).  In addition to this mode, $D_s(4430)\to D^{(*)}K^{(*)}$ are also possible where $D^{(*)}$ and $K^{(*)}$ are excited states of $D$ and $K$, respectively, if they are kinematically allowed. These channels will be identified with $D+K$ production accompanying one or more pions and they are also signal channels for the $Z^+(4430)$.  However, among these channels, $D_s(4430)\to DK$ will be dominant because of the phase space effect.

\begin{figure}[t]
\begin{center}
\begin{picture}(270,110)
\ArrowLine(100,67)(20,67)
\ArrowLine(180,94)(100,67)
\ArrowLine(120,60)(180,80)
\Text(20,75)[l]{\large $\bar{d}$}
\Text(20,44)[l]{\large $b$}
\Text(180,102)[r]{\large $\bar{u}$}
\Text(180,72)[r]{\large $s$}
\Photon(100,52)(100,67){2}{3}
\Text(88,60)[]{$W$}
\Gluon(120,60)(110,70){2}{3}
\ArrowLine(20,52)(100,52)
\ArrowLine(100,52)(240,10)
\ArrowLine(240,24)(120,60)
\Text(240,32)[r]{\large $\bar{c}$}
\Text(240,4)[r]{\large $c$}
\Photon(180,42)(210,50){3}{3}
\Text(195,55)[]{$W$}
\ArrowLine(210,50)(240,50)
\ArrowLine(240,65)(210,50)
\Text(250,67)[r]{\large $\bar{d}$}
\Text(250,50)[r]{\large $u$}
\Oval(150,45)(20,8)(327)
\Text(157,56)[]{\large $\bar{s}$}
\Text(142,33)[]{\large $c$}
\Text(0,60)[]{\red \large $\bar B^0$}
\Text(198,90)[]{\red \large $K^-$}
\Text(265,60)[]{\red \large $\pi^+$}
\Text(255,15)[]{\red \large $\psi'$}
\Text(132,20)[]{\red \large $D_s$}
\end{picture}
\caption{Decay process through $D_s$: $\bar B^0 \to Z^+(4430)K^- \to \psi' \pi^+ K^-$.}
\label{decay_Ds0} 
\end{center}
\end{figure}
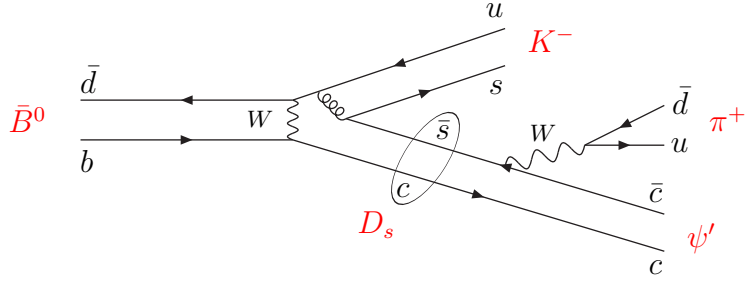
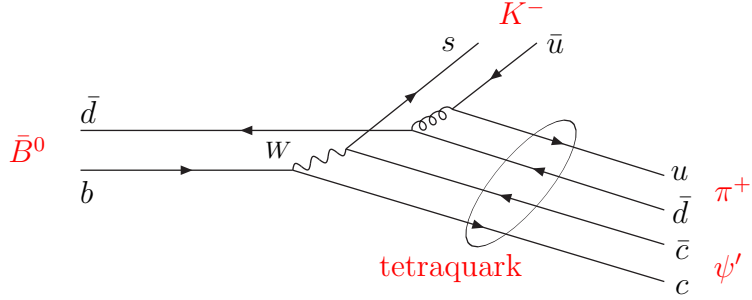
\begin{figure}[t]
\begin{center}
\begin{picture}(270,115)
\ArrowLine(145,67)(20,67)
\ArrowLine(240,37)(145,67)
\ArrowLine(120,60)(170,100)
\Text(20,75)[l]{\large $\bar{d}$}
\Text(20,44)[l]{\large $b$}
\Text(162,100)[r]{\large $s$}
\Text(203,100)[r]{\large $\bar{u}$}
\Gluon(145,67)(160,75){2}{3}
\Text(95,60)[]{$W$}
\Photon(100,52)(120,60){2}{3}
\ArrowLine(20,52)(100,52)
\ArrowLine(100,52)(240,10)
\ArrowLine(240,24)(120,60)
\Text(250,22)[r]{\large $\bar{c}$}
\Text(250,8)[r]{\large $c$}
\ArrowLine(192,100)(160,75)
\ArrowLine(160,75)(240,50)
\Text(250,37)[r]{\large $\bar{d}$}
\Text(250,51)[r]{\large $u$}
\Oval(186,47)(30,10)(320)
\Text(0,60)[]{\red \large $\bar B^0$}
\Text(187,112)[]{\red \large $K^-$}
\Text(267,45)[]{\red \large $\pi^+$}
\Text(265,15)[]{\red \large $\psi'$}
\Text(160,15)[]{\red \large tetraquark}
\end{picture}
\caption{Decay process through tetraquark: $\bar B^0 \to Z^+(4430)K^- \to \psi' \pi^+ K^-$.}
\label{decay_Tetra0} 
\end{center}
\end{figure}

\begin{figure}[t]
\begin{center}
\begin{picture}(390,260)(0,-20)
\ArrowLine(30,216)(90,216)
\ArrowLine(95,200)(30,200)
\Text(20,218)[]{$c$}
\Text(20,200)[]{$\bar{s}$}
\ArrowLine(90,216)(130,245)
\ArrowLine(140,233)(95,200)
\Text(135,250)[]{$c$}
\Text(145,238)[]{$\bar{c}$}
\ArrowLine(150,170)(125,188)
\ArrowLine(125,188)(156,181)
\Text(157,167)[]{$\bar{d}$}
\Text(165,179)[]{$u$}
\Photon(95,200)(125,188){3}{3.5}
\Text(110,175)[]{$W$}
\Text(0,208)[]{\large $D_{s}^+$}
\Text(157,253)[]{\large $\psi'$}
\Text(180,166)[]{\large $\pi^+$}
\Text(80,140)[]{\large (a)}
\ArrowLine(240,216)(300,216)
\ArrowLine(300,200)(240,200)
\Text(230,218)[]{$c$}
\Text(230,200)[]{$\bar{s}$}
\ArrowLine(300,216)(340,245)
\ArrowLine(350,233)(320,208)
\Text(346,250)[]{$c$}
\Text(357,238)[]{$\bar{d}$}
\ArrowLine(340,171)(300,200)
\ArrowLine(320,208)(350,183)
\Text(346,167)[]{$\bar{c}$}
\Text(356,179)[]{$u$}
\Photon(300,200)(320,208){3}{2}
\Text(310,213)[]{$W$}
\Text(210,208)[]{\large $D_{s}^+$}
\Text(367,253)[]{\large $D^+$}
\Text(367,163)[]{\large $\bar{D}^0$}
\Text(290,140)[]{\large (b)}
\ArrowLine(30,66)(90,66)
\ArrowLine(90,50)(30,50)
\Text(20,68)[]{$c$}
\Text(20,50)[]{$\bar{s}$}
\ArrowLine(90,66)(130,95)
\ArrowLine(140,85)(105,58)
\Text(136,100)[]{$c$}
\Text(147,88)[]{$\bar{u}$}
\ArrowLine(130,21)(90,50)
\ArrowLine(105,58)(140,31)
\Text(136,17)[]{$\bar{s}$}
\Text(147,28)[]{$u$}
\Text(0,58)[]{\large $D_{s}^+$}
\Text(157,103)[]{\large $D^0$}
\Text(157,13)[]{\large $K^+$}
\Text(80,-10)[]{\large (c)}
\ArrowLine(240,66)(300,66)
\ArrowLine(300,50)(240,50)
\Text(230,68)[]{$c$}
\Text(230,50)[]{$\bar{s}$}
\ArrowLine(300,66)(340,95)
\ArrowLine(350,85)(315,58)
\Text(346,100)[]{$c$}
\Text(357,88)[]{$\bar{d}$}
\ArrowLine(340,21)(300,50)
\ArrowLine(315,58)(350,31)
\Text(346,17)[]{$\bar{s}$}
\Text(357,28)[]{$d$}
\Text(210,58)[]{\large $D_{s}^+$}
\Text(367,103)[]{\large $D^+$}
\Text(367,13)[]{\large $K^0$}
\Text(290,-10)[]{\large (d)}
\end{picture}
\caption{$D_{s}$ Decay}
\label{decay_Ds} 
\end{center}
\end{figure}
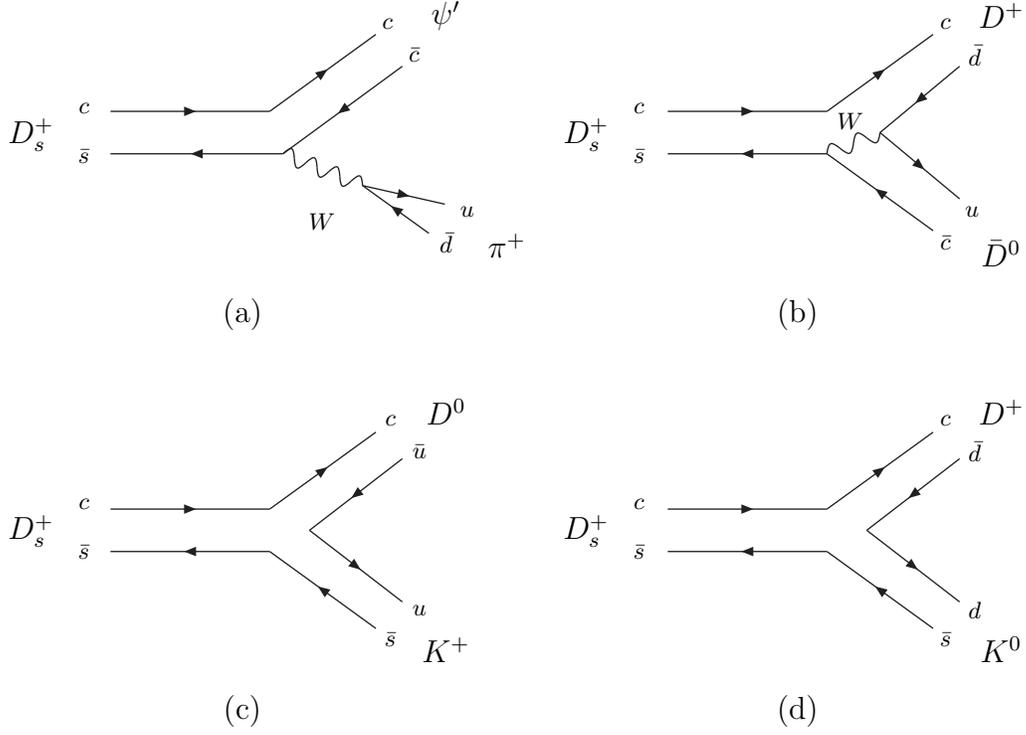
\begin{figure}[t]
\begin{center}
\begin{picture}(390,220)(0,-10)
\ArrowLine(30,200)(90,200)
\ArrowLine(90,190)(30,190)
\ArrowLine(30,180)(90,180)
\ArrowLine(90,170)(30,170)
\Text(20,200)[]{$c$}
\Text(20,190)[]{$\bar{c}$}
\Text(20,180)[]{$u$}
\Text(20,170)[]{$\bar{d}$}
\ArrowLine(90,200)(140,210)
\ArrowLine(140,200)(90,190)
\Text(150,213)[]{$c$}
\Text(150,202)[]{$\bar{c}$}
\ArrowLine(90,180)(140,170)
\ArrowLine(140,160)(90,170)
\Text(150,170)[]{$u$}
\Text(150,159)[]{$\bar{d}$}
\Text(0,185)[]{\large $T^+$}
\Text(170,210)[]{\large $\psi'$}
\Text(170,165)[]{\large $\pi^+$}
\Text(90,130)[]{\large (a)}
\ArrowLine(240,200)(300,200)
\ArrowLine(300,190)(240,190)
\ArrowLine(240,180)(300,180)
\ArrowLine(300,170)(240,170)
\Text(230,200)[]{$c$}
\Text(230,190)[]{$\bar{d}$}
\Text(230,180)[]{$u$}
\Text(230,170)[]{$\bar{c}$}
\ArrowLine(300,200)(350,210)
\ArrowLine(350,200)(300,190)
\Text(360,213)[]{$c$}
\Text(360,202)[]{$\bar{d}$}
\ArrowLine(300,180)(350,170)
\ArrowLine(350,160)(300,170)
\Text(360,170)[]{$u$}
\Text(360,159)[]{$\bar{c}$}
\Text(210,185)[]{\large $T^+$}
\Text(380,210)[]{\large $D^+$}
\Text(380,165)[]{\large $\bar{D}^0$}
\Text(300,130)[]{\large (b)}
\ArrowLine(30,70)(90,70)
\ArrowLine(70,55)(30,60)
\ArrowLine(30,50)(70,55)
\ArrowLine(90,40)(30,40)
\Text(20,70)[]{$c$}
\Text(20,60)[]{$\bar{d}$}
\Text(20,50)[]{$u$}
\Text(20,40)[]{$\bar{c}$}
\ArrowLine(90,70)(140,80)
\ArrowLine(140,70)(105,55)
\Text(150,83)[]{$c$}
\Text(150,72)[]{$\bar{u}$}
\ArrowLine(105,55)(140,40)
\ArrowLine(140,30)(90,40)
\Text(150,40)[]{$u$}
\Text(150,29)[]{$\bar{s}$}
\Photon(70,55)(90,40){3}{2.5}
\Text(90,55)[]{$W$}
\Text(0,55)[]{\large $T^+$}
\Text(170,80)[]{\large $D^0$}
\Text(170,35)[]{\large $K^+$}
\Text(90,0)[]{\large (c)}
\ArrowLine(240,70)(300,70)
\ArrowLine(280,55)(240,60)
\ArrowLine(240,50)(280,55)
\ArrowLine(300,40)(240,40)
\Text(230,70)[]{$c$}
\Text(230,60)[]{$\bar{d}$}
\Text(230,50)[]{$u$}
\Text(230,40)[]{$\bar{c}$}
\ArrowLine(300,70)(350,80)
\ArrowLine(350,70)(315,55)
\Text(360,83)[]{$c$}
\Text(360,72)[]{$\bar{d}$}
\ArrowLine(315,55)(350,40)
\ArrowLine(350,30)(300,40)
\Text(360,40)[]{$d$}
\Text(360,29)[]{$\bar{s}$}
\Photon(280,55)(300,40){3}{2.5}
\Text(300,55)[]{$W$}
\Text(210,55)[]{\large $T^+$}
\Text(380,80)[]{\large $D^+$}
\Text(380,35)[]{\large $K^0$}
\Text(300,0)[]{\large (d)}
\end{picture}
\caption{Tetraquark Decay}
\label{decay_Tetra} 
\end{center}
\end{figure}
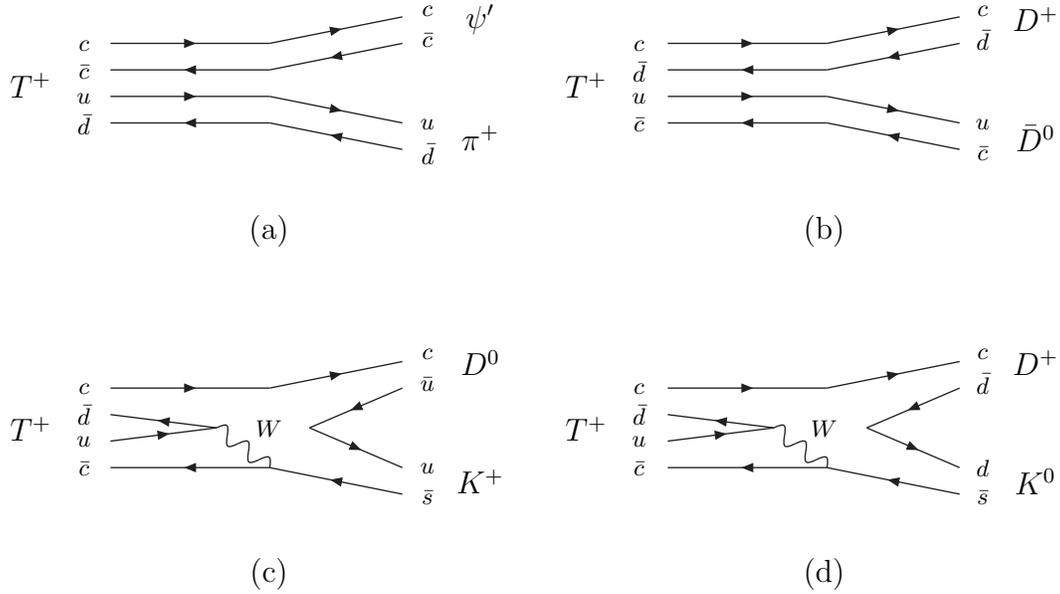

\begin{enumerate}
\item Figure 1 in the case of $Z={\rm radially~excited}~D_s$ involves two weak
bosons, hence this is the second order process in the weak interaction.  On the other hand,
Fig. 2 in the case of $Z={\rm tetraquark}$ involves only one weak boson and thus this is the first order weak interaction process.  However the latter case needs to be multiplied with a recombination 
probability to form a tetraquark, which must be ${\cal O}(10^{-5})$ as mentioned early. Thus, it is expected that these two cases would be in the same order of magnitude for a chain decay rate. 
Therefore, we cannot distinguish a tetraquark and a radially excited $D_s$ as long as we look at the product of
branching ratios alone given by Eq. (\ref{prdrate}).
\item As one can easily notice from Fig. 2, three charged states, $\pm$ and $0$, of $Z(4430)$ must be available if $Z$ is a tetraquark state because $[c\bar cu\bar d]$,
$[c\bar cd\bar u]$, $[c\bar c(u\bar u-d\bar d)/\sqrt{2}]$, and $[c\bar c(u\bar u+d\bar d)/\sqrt{2}]$ are possible. On the other hand,
if $Z$ is a radial excitation of $D_s$, then only the $\pm$ charged states are possible
because only $c\bar s$ and $\bar c s$ are possible states. Namely if $Z$ is a tetraquark state, its isospin
is one and/or singlet, while if $Z$ is a radially excited state of $D_s$, then its isospin is $1/2$.
\item To distinguish these two possibilities, i.e., the case of a tetraquark and the case of a radially excited
$D_s$, let us consider the diagrams, Figs. 3 and
4, for $Z=D_s(4430)$ and $Z=T(4430)$ ($T$ means a 
tetraquark), respectively.  If we assume $Z=D_s(4430)$, then the process
$D_s^+(4430) \to \psi' \pi^+$ (Fig. 3(a)) goes through one weak boson while the
processes $D_s^+(4430) \to D^0 K^+$(Fig. 3(c))$/D^{+}K^{0}$(Fig. 3(d)) 
are strong decays. The process $D_s^+(4430) \to \bar D^0 D^+$ (Fig. 3(b)) needs to
go through one weak boson exchange, hence the following ratios are expected to be obtained.
\begin{equation}
  \frac{\Gamma\left(D_s^+(4430)\to D^+K^0/D^0K^+\right)}{\Gamma\left(D_s^+(4430) \to \psi' \pi^+
  \right)} \gg 1, \quad
  \frac{\Gamma\left(D_s^+(4430)\to \bar D^0D^+\right)}{\Gamma\left(D_s^+(4430) \to \psi'\pi^+
  \right)} \approx 1 \label{Dratio} .
\end{equation}
On the other hand, if we assume $Z=T(4430)$, either the process
$T^+(4430) \to D^0K^+$ (Fig. 4(c)) or $T^+(4430) \to D^+K^0$
(Fig. 4(d)) goes through one weak boson exchange. The process
$T^+(4430) \to \bar D^0D^+$ (Fig. 4(b)) is not a weak 
interaction but a strong interaction process. Hence the following ratios between these processes 
and the process 
$T^+(4430) \to \psi' \pi^+$ (Fig. 4(a)) are expected to be obtained.
\begin{equation}
  \frac{\Gamma\left(T^+(4430)\to D^+K^0/D^0K^+\right)}{\Gamma\left(T^+(4430) \to \psi' \pi^+
  \right)} \ll 1, \quad
  \frac{\Gamma\left(T^+(4430)\to \bar D^0D^+\right)}{\Gamma\left(T^+(4430) \to \psi'\pi^+
  \right)} \approx 1 \label{Tratio} .
\end{equation}
Therefore if one measures the process $Z\to DK$ and its ratios to 
$Z \to \psi' \pi$, i.e., Eqs. (\ref{Dratio}) and (\ref{Tratio}), then one can 
distinguish these two possibilities whether $Z$ is $D_s(4430)$ or $T(4430)$. 

\item Since a radially excited $D_s$ decays to $\psi'+\pi$ via weak interactions, the partial decay width $\Gamma(D_s^+(4430)\to \psi' \pi)$ would be very small, contrary to the one for $Z^{+}(4430)=T^{+}(4430)$, $\Gamma(T^{+}(4430)\to \psi'\pi)$, which is via a strong interaction decay.  However, as shown in Eq. (\ref{Dratio}), $D_s^+(4430)$ will dominantly decay to $D^{+}K^{0}/D^{0}K^{+}$ and hence the total decay width of $D_s^+(4430)$ becomes an order of
a strong decay.  As described in the beginning, in the Belle experiment only the total decay width
of Eq. (\ref{masswidth}) being in the order of
strong interactions, is obtained by fitting the invariant mass of $\psi'+\pi$ to the $S$-wave Breit-Wigner resonance formula.  Therefore, the present experiment does not rule out the possibility of $Z^{+}(4430)$ being the $D_s^{+}(4430)$.
\end{enumerate}
The Belle Collaboration discovered the rather narrow resonance $Z^{+}(4430)$ 
in the invariant mass distribution of $\psi' +\pi^{+}$ in the decay mode 
$\bar B^{0}\to K^{-}\psi'\pi^{+}$.  However, in the $\bar B^{0}$ decay there
are other decay modes, $\bar B^{0}\to K^{-}K^{0}D^{+}/K^{-}K^{+}D^{0}$ or 
$\bar B^{0}\to K^{-}\bar D^{0}D^{+}$, in addition to this special mode 
$\bar B^{0}\to K^{-}Z^{+}(4430) \to K^{-}\psi'\pi^{+}$.  If the $Z^{+}(4430)$ 
is a radially excited $D_s$ state, one can find $Z^{+}(4430)$ more frequently 
in the invariant distribution of $K^{0}D^{+}/K^{+}D^{0}$ in the mode 
$\bar B^{0}\to K^{-}K^{0}D^{+}/K^{-}K^{+}D^{0}$.  On the other hand, if the 
$Z^{+}(4430)$ is a tetraquark, it must be very difficult to find 
it in the mode 
$\bar B^{0}\to K^{-}K^{0}D^{+}/K^{-}K^{+}D^{0}$. Therefore, this type of decay mode is very 
important to determine whether $Z^{+}(4430)$ is a tetraquark or a 
radially excited $D_s$. 

Recently the CDF Collaboration reported the observation of the decay mode 
$B_c^\pm\to J/\psi \pi^\pm$ \cite{CDF06}.  This must become a very interesting
example to test our model: in this decay mode the heavy
mesons in the initial and final states have both zero nodes and therefore, 
this decay rate should be very large compared with the rate for the process
$B_c^\pm\to \psi' \pi^\pm$ with the node of $\psi'$ to be 1, which might be 
strongly suppressed or not be observed.

In this letter, we have discussed the possible interpretation of $Z^{+}(4430)$
 as a radially excited $D_s$ state. We presented the mass levels of radially 
excited $D_s$ states and gave the 
reasonable explanation on why the mode $Z\to J/\psi \pi$ was not observed 
in the Belle experiment.  As shown in Table \ref{masslevel}, we can see a wealth of excited $D_s$ states.  We would like to stress that the experimental search for resonances in $DK$ and $D^{(*)}K^{(*)}$ invariant mass distributions is very interesting not only for observing $Z^+(4430)$ but also for discovery of $n=2$ and $n=3$ excited $D_s$ states.  Related to this, it is remarkable that the $n=2$ $^3S_1(1^-)$ 2715 MeV and $n=2$ $^3P_0(0^+)$ 2856 MeV were already observed and reproduced well by our semirelativistic model, as shown in Table 1.

  We would like to urge the analysis on the invariant mass 
distribution of $K^{0}D^{+}/K^{+}D^{0}$ in the decay mode 
$\bar B^{0}\to K^{-}K^{0}D^{+}/K^{-}K^{+}D^{0}$ which must contain a fruitful
physics of excited $D_s$ states. 
\medskip

\begin{center}
Acknowledgments
\end{center}
\medskip

We would like to thank S. Aoki for helping us in understanding the experimental data.


\def\Journal#1#2#3#4{{#1} {\bf #2} (#4), #3}
\def\NIM{Nucl. Instrum. Methods}
\def\NIMA{Nucl. Instrum. Methods A}
\def\NPB{Nucl. Phys. B}
\def\PLB{Phys. Lett. B}
\def\PRL{Phys. Rev. Lett.}
\def\PRD{Phys. Rev. D}
\def\PRO{Phys. Rev.}
\def\PTP{Prog. Theor. Phys.}
\def\ZPC{Z. Phys. C}
\def\EPJC{Eur. Phys. J. C}
\def\EPJA{Eur. Phys. J. A}
\def\PR{Phys. Rept.}
\def\IJM{Int. J. Mod. Phys. A}

\end{document}